\documentclass[12 pt, amsfonts, amssymb,color]{article}

\usepackage{amsmath,amssymb,amsfonts,latexsym,float,graphics,epsfig}

\usepackage{epstopdf}%

\begin{document}

\renewcommand\theequation{\arabic{section}.\arabic{equation}}
\catcode`@=11 \@addtoreset{equation}{section}
\newtheorem{lem}{Lemma}[section]
\newtheorem{cor}{Corollary}[section]
\newcommand{\be}{\begin{equation}}
\newcommand{\ee}{\end{equation}}

\newcommand{\equal}{\!\!\!&=&\!\!\!}
\newcommand{\rd}{\partial}
\newcommand{\g}{\hat {\cal G}}
\newcommand{\bo}{\bigodot}
\newcommand{\res}{\mathop{\mbox{\rm res}}}
\newcommand{\diag}{\mathop{\mbox{\rm diag}}}
\newcommand{\Tr}{\mathop{\mbox{\rm Tr}}}
\newcommand{\const}{\mbox{\rm const.}\;}
\newcommand{\cA}{{\cal A}}
\newcommand{\bA}{{\bf A}}
\newcommand{\Abar}{{\bar{A}}}
\newcommand{\cAbar}{{\bar{\cA}}}
\newcommand{\bAbar}{{\bar{\bA}}}
\newcommand{\cB}{{\cal B}}
\newcommand{\bB}{{\bf B}}
\newcommand{\Bbar}{{\bar{B}}}
\newcommand{\cBbar}{{\bar{\cB}}}
\newcommand{\bBbar}{{\bar{\bB}}}
\newcommand{\bC}{{\bf C}}
\newcommand{\cbar}{{\bar{c}}}
\newcommand{\Cbar}{{\bar{C}}}
\newcommand{\Hbar}{{\bar{H}}}
\newcommand{\cL}{{\cal L}}
\newcommand{\bL}{{\bf L}}
\newcommand{\Lbar}{{\bar{L}}}
\newcommand{\cLbar}{{\bar{\cL}}}
\newcommand{\bLbar}{{\bar{\bL}}}
\newcommand{\cM}{{\cal M}}
\newcommand{\bM}{{\bf M}}
\newcommand{\Mbar}{{\bar{M}}}
\newcommand{\cMbar}{{\bar{\cM}}}
\newcommand{\bMbar}{{\bar{\bM}}}
\newcommand{\cP}{{\cal P}}
\newcommand{\cQ}{{\cal Q}}
\newcommand{\bU}{{\bf U}}
\newcommand{\bR}{{\bf R}}
\newcommand{\cW}{{\cal W}}
\newcommand{\bW}{{\bf W}}
\newcommand{\bZ}{{\bf Z}}
\newcommand{\Wbar}{{\bar{W}}}
\newcommand{\Xbar}{{\bar{X}}}
\newcommand{\cWbar}{{\bar{\cW}}}
\newcommand{\bWbar}{{\bar{\bW}}}
\newcommand{\abar}{{\bar{a}}}
\newcommand{\nbar}{{\bar{n}}}
\newcommand{\pbar}{{\bar{p}}}
\newcommand{\tbar}{{\bar{t}}}
\newcommand{\ubar}{{\bar{u}}}
\newcommand{\utilde}{\tilde{u}}
\newcommand{\vbar}{{\bar{v}}}
\newcommand{\wbar}{{\bar{w}}}
\newcommand{\phibar}{{\bar{\phi}}}
\newcommand{\Psibar}{{\bar{\Psi}}}
\newcommand{\bLambda}{{\bf \Lambda}}
\newcommand{\bDelta}{{\bf \Delta}}
\newcommand{\p}{\partial}
\newcommand{\om}{{\Omega \cal G}}
\newcommand{\ID}{{\mathbb{D}}}
\newcommand{\pr}{{\prime}}
\newcommand{\prr}{{\prime\prime}}
\newcommand{\prrr}{{\prime\prime\prime}}

\title{Noetherian symmetries of noncentral forces with drag term\\
}
\author{A Ghose-Choudhury\thanks{%
E-mail: aghosechoudhury@gmail.com} \\
Department of Physics, Surendranath College,\\
24/2 Mahatma Gandhi Road, Calcutta 700009, India\\
\\
Partha Guha\thanks{%
E-mail: partha@bose.res.in}\\
SN Bose National Centre for Basic Sciences \\
JD Block, Sector III, Salt Lake,
Kolkata 700106, India \\
\and Andronikos Paliathanasis\thanks{%
E-mail: anpaliat@phys.uoa.gr} \\
Instituto de Ciencias F\'{\i}sicas y Matem\'{a}ticas,\\ Universidad Austral de
Chile, Valdivia, Chile\\
\and PGL Leach\thanks{%
E-mail: leach@ucy.ac.cy}\thanks{%
School of Mathematics, Statistics and Computer Science, University of
KwaZulu-Natal, Private Bag X54001, Durban 4000, Republic of South Africa and
Department of Mathematics, Durban University of Technology, PO Box 1334,
Durban 4000, Republic of South Africa} \\
Department of Mathematics and Statistics, \\
University of Cyprus, Lefkosia 1678, Cyprus. \\
}
\date{ }
\maketitle
\vspace{.05in}
\hspace{1.10in}

\begin{center}
{\it Dedicated to Sir Michael Berry on his 75th birthday with great respect and admiration}
\end{center}

\begin{abstract}
\textit{We consider the Noetherian symmetries of second-order ODEs subjected to forces
with nonzero curl. Both position and velocity dependent forces are considered.
In the former case the first integrals are shown to follow from the symmetries
of the celebrated Emden-Fowler equation.}
\end{abstract}


\renewcommand\theequation{\arabic{section}.\arabic{equation}} \catcode`@=11 %
\@addtoreset{equation}{section}
\newtheorem{axiom}{Definition}[section] %
\newtheorem{theorem}{Theorem}[section] \newtheorem{axiom2}{Example}[section] %
\newtheorem{axiom3}{Lemma}[section] \newtheorem{prop}{Proposition}[section] %
\newtheorem{axiom5}{Corollary}[section]

\smallskip

\paragraph{Mathematics Classification (2010)}

: 34C14, 34C20.

\smallskip

\paragraph{Keywords:}

Curl forces, anisotropic central forces, Emden-Fowler equation,

\section{Introduction}

The motion of a particle under the influence of a central force is a
standard topic of Classical Mechanics and is treated extensively in almost
all text books on the subject. The radial nature of the force implies the
conservation of angular momentum and greatly simplifies the analysis of the
radial equation, with the orbit being determined by Binet's formula.
However, in most cases the possibility of the central forces being of
anisotropic character is usually not treated. \newline
Newtonian forces depending on position and having a nonvanishing curl are usually termed as curl forces.
In \cite{BS}it was shown that for a plananr isotropic force, $F=f(r){\mathbf{e}}_\theta$
when $f(r)=r^\mu$, with a nonvanishing curl one can quite generally map the radial equation to the
Emden-Fowler equation by defining a transformation of both the independent
and dependent variables.

Incidentally the issue of such forces did
not escape Whittaker's \cite{Whittaker 44 a} attention. In his book\footnote{%
Article 52 of 4th edition (1937), page 96} the problem of the most general
field of force under which a given curve can be described is treated.
Starting with a curve $\phi(x,y) = c$ the general form of the component of
the acceleration $(X,Y)$ are derived. This expression involves an arbitrary
function $u$ of $x,y$, which is related to the square of the velocity of
particle. In general the curl condition is $\partial_xY - \partial_yX \neq 0$.
Of course there is no mention of the possibility of deriving Hamiltonians
in this context. In recent times a fairly general treatment of the Ermakov
\cite{CA} and generalised Ermakov system \cite{HG} was also performed which
treatment involved forces depending upon both $r$ and $\theta$ with a
nonvanishing curl\footnote{%
A somewhat similar analysis is found in \cite{GL}, but there the possibility
that the curl of the force be nonzero was not stressed.}. These analyses
were motivated mainly by a desire to examine if the equation was
linearisable. Berry and Shukla \cite{BS1} showed that the force on a
particle with complex electric polarizability is known to be not derivable
from a potential, i.e., its curl is nonzero. In general curl forces are
 Newtonian but not Hamiltonian or Lagrangian. However, recently the
Hamiltonian formalism of curl forces has been studied in \cite{BS2}.\newline

\smallskip

In \cite{BS} the authors performed a detailed analysis of the nature of the
motion for specific values of the exponent $\mu$ of the force. In particular
their analysis of the case $\mu = -1$ is interesting and begs the question
of the possible implications of such a motion evoking as it does an uncanny
similarity with the Aharonov-Bohm effect. Furthermore in course of their
analysis they determined two first integrals of motion corresponding to $%
\mu=-3/2$ and $\mu=-5/3$, respectively. The Emden-Fowler (EF) equation which
forms the cornerstone of their work is a well-known nonlinear ordinary
differential equation and has been extensively studied by Leach \textit{et al%
} \cite{PL1, PL2, PL3} from the point of view of its symmetries.
Consequently one of the objectives of the present article is to show that
these first integrals are actually the fruits of the existence of a
variational (Noetherian) symmetry of the Emden-Fowler equation.\newline

\smallskip

Consider a Lagrangian system $ L : {\Bbb R} \times TQ \to {\Bbb R}$, on a configuration space $Q$ with local coordinates $q = (q_1, \cdots ,q_n)$.
The action of an one-parameter group of diffeomorphisms $\Phi_s$ on ${\Bbb R} \times Q$ with the
induced vector field reads
$$
{\cal X} = \eta(t,q)\frac{\partial}{\partial t} + \sum_{i}( \xi^i(t,q)\frac{\partial}{\partial q_i}
+ \hat{\xi}^i(t,q)\frac{\partial}{\partial {\dot q}_i}),
$$
where $$ \hat{\xi}^i(t,q) = \frac{\partial \xi^i}{\partial t} - {\dot q}_i\frac{\partial \eta}{\partial t} +
\sum_{j}  (\frac{\partial \xi^i}{\partial q_j} {\dot q}_j -  {\dot q}_i\frac{\partial \eta}{\partial \partial q_j} {\dot q}_j).
$$
The group $\Phi_s$ is a Noetherian symmetry of the Lagrangian system if it preserves
the action functional $S = \int_{a}^{b} L(t,q,\dot{q}) dt$, if
\begin{equation}
\frac{\partial L}{\partial t}\eta + \sum_{i} (\xi^i(t,q)\frac{\partial L}{\partial q_i} + \hat{\xi}^i(t,q)\frac{\partial L}{\partial {\dot q}_i})
+ L(\frac{\partial \eta}{\partial t} + \frac{\partial \eta}{\partial q_j}{\dot q}_j )  = 0
\end{equation}

The Noether theorem states that if $\Phi_s$ is a Noether symmetry then
\begin{equation}
{\Bbb J}(t,q, {\dot q}) = \sum_i \frac{\partial L}{\partial {\dot q}_i} ( \xi^i - \eta {\dot q}_i) + L\eta.
\end{equation}

\smallskip

There are several possible ways to generalise the Berry-Shukla construction.
In this paper we briefly outline the curl force in the presence of a
coordinate-dependent dissipative force and show that the system can be
mapped to the Lane-Emden equation, which appears in the study of stellar
structure. Polytropes are a family of equations of state for which the
pressure $P$ is given as a power of density $\rho$, $P = \kappa \rho^{\gamma}
$, where $\kappa$ and $\gamma$ are constants. The Lane-Emden equation
\begin{equation*}
\frac{1}{\xi^2}\frac{d}{d\xi}\left(\xi^2 \frac{d\theta}{d\xi}\right) +
\theta^n = 0
\end{equation*}
combines this $P$ and $\rho$ relation and the equation of hydrostatic
equilibrium. This was originally proposed by Jonathan Lane \cite{Lane} and
was analysed by Emden \cite{Emden}. The Lane-Emden equation can be solved
analytically only for a few special, integer values of the index $n: 0, 1$
and $5$ \cite{Mach} and for all other values of $n$ we must resort to
numerical solutions. Several applications of the Emden-Fowler and Lane-Emden
equations of various forms arising in astrophysics \cite{Chandra} and
nonlinear dynamics have been reported. The reader is also referred to the
now oldish paper by Wang \cite{Wang 76 a} for a sampling of citations of
papers dealing with the equation using various approaches.

\smallskip

The paper is organized as follows. In Section 2 we introduce nonisotropic
curl forces and briefly recollect the results of Athorne \cite{CA} and Haas and
Goedert \cite{HG}. In Section 3 we show that from the particular solution of the EF
equation it is possible to arrive at the results of \cite{BS}. Finally by
introducing the Lagrangian of the Emden-Fowler equation and taking into
consideration its symmetries one can easily derive the first integrals
stated in Berry's and Shukla's paper.
\section{Motion under noncentral forces}

The motion of a point particle in the plane, taken for convenience to be of
unit mass, is best studied in terms of polar coordinates, $r$ and $\theta$,
in terms of which the components of the equation of motion are
\begin{align}  \label{e2}
\ddot{r}-r\dot{\theta}^2&=F_r(r, \theta), \\
r\ddot{\theta}+2\dot{r}\dot{\theta}&=F_\theta(r,\theta).
\end{align}
In plane polar coordinates the condition $\nabla\times F\ne 0$ translates
into the requirement that
\begin{equation}  \label{e3}
\frac{\partial}{\partial r}\left(rF_{\theta}\right)\ne \frac{\partial F_r}{%
\partial\theta}.
\end{equation}
For the generalised Ermakov system of \cite{HG} the radial and transverse
components of the force are given by
\begin{align}  \label{e5}
F_r(r, \theta)&=-w^2 r+\frac{1}{r^3}U(\theta), \\
F_\theta(r, \theta) &=-\frac{1}{r^3}\frac{dV}{d\theta}.
\end{align}
Here $U(\theta)$ and $V(\theta)$ are arbitrary functions and
one may verify that the force satisfies (\ref{e3}) and is therefore a
curl force. For the generalised Ermakov system there exists the well-known
Lewis-Ray-Reid (LRR) invariant
\begin{equation}  \label{e6}
I=\frac{1}{2}(r^2\dot{\theta})^2 +V(\theta).
\end{equation}

{In this connection mention has to be made of the method used by Gorringe
and Leach \cite{GL} to deduce a first integral for a planar system governed
by a noncentral force the radial and tangential components of which are
given by
\begin{equation*}
F_r(r, \theta)=-\left[\frac{U^{\prime\prime}(\theta)+U(\theta)}{r^2}+\frac{%
2V^\prime(\theta)}{r^{3/2}}\right],\;\; F_\theta(r,\theta)=-\frac{V(\theta)}{%
r^{3/2}}.
\end{equation*}
Here $U(\theta)$ and $V(\theta)$ are arbitrary functions. Note that for such
a force $\nabla\times F\ne 0$ if and only if $U(\theta)$ and $V(\theta)$ are not  sinusoidal
functions of their arguments.}

\begin{figure}
\centering
\includegraphics[scale=0.5]{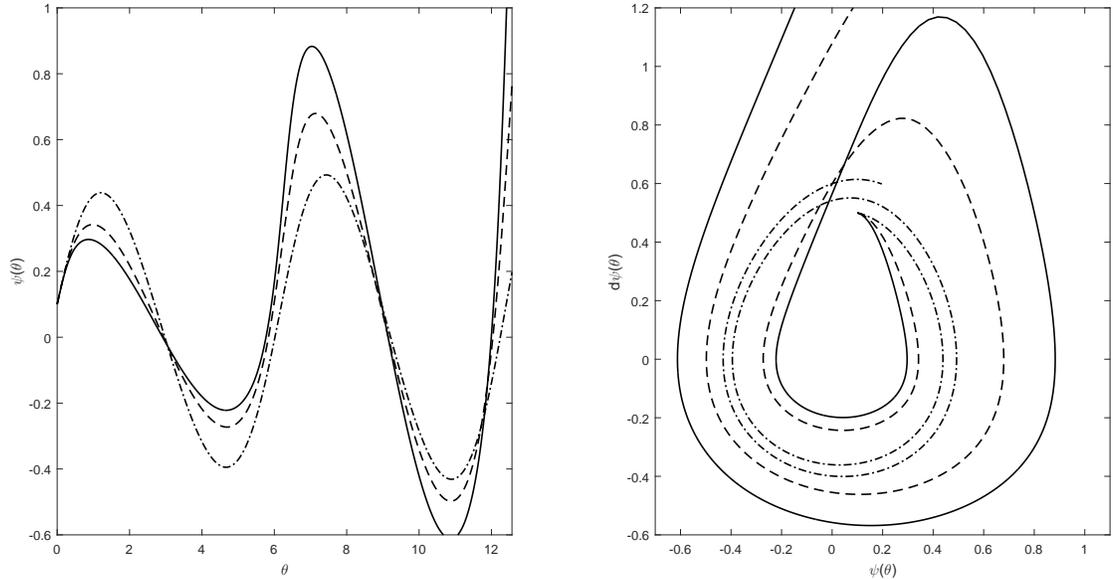}
\caption{Numerical solution of equation \ref{e13}) for different values of the constant of motion I greater than one. Solid line is for $I=1.1$, dash-dash line is for $I=1.2$, while dash-dot line is for $I=2$. Left figure is the $\theta-\psi$ diagram while right figure is the phase portrait $\psi-d\psi$}
\label{fig1}
\end{figure}

\begin{figure}
\centering
\includegraphics[scale=0.5]{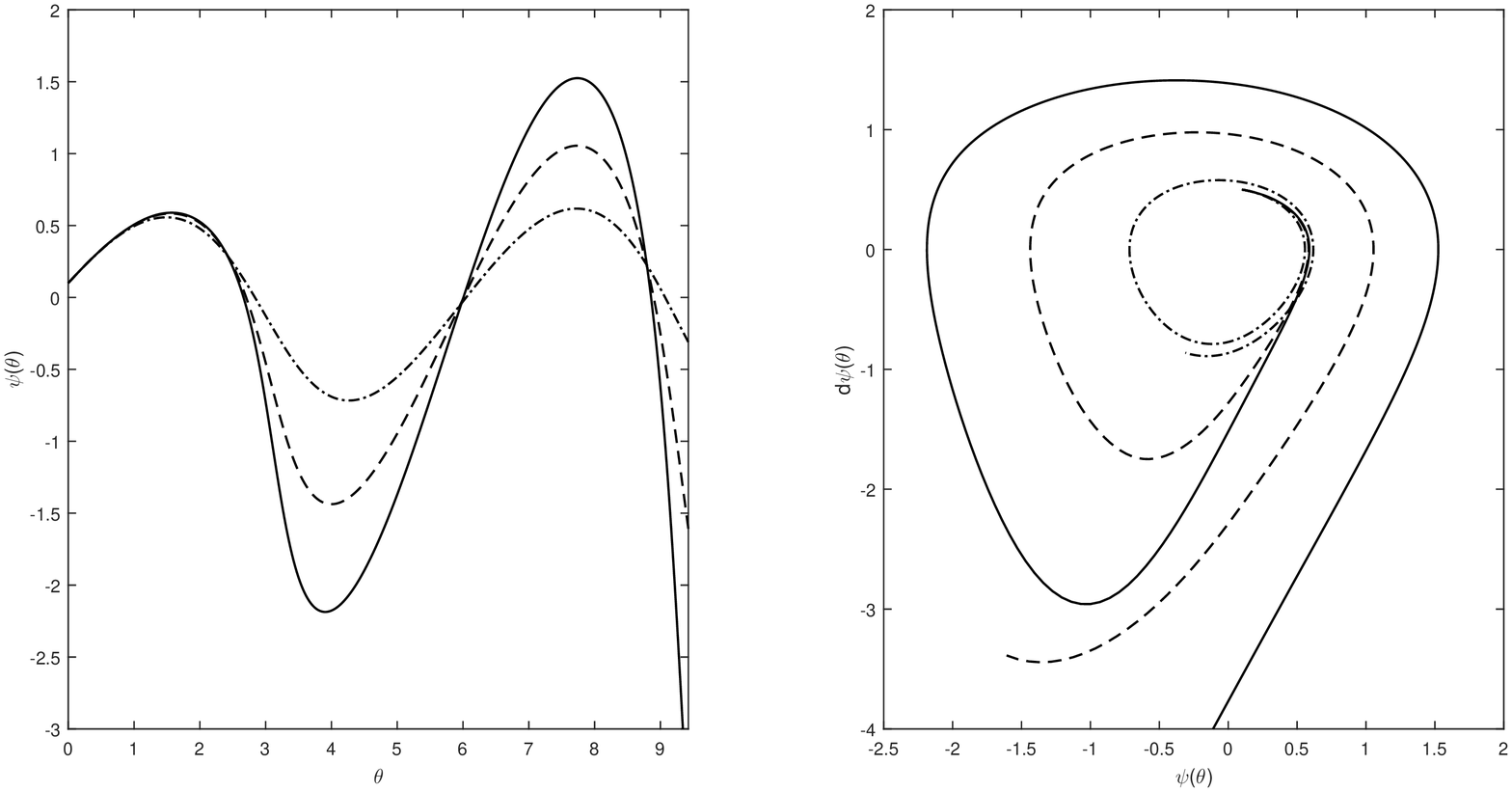}
\caption{Numerical solution of equation (\ref{e13}) for different values of the constant of motion I, smaller than minus one. Solid line is for $I=1.1$, dash-dash line is for $I=-1.2$, while dash-dot line is for $I=-2$. Left figure is the $\theta-\psi$ diagram while right figure is the phase portrait $\psi-d\psi$}
\label{fig2}
\end{figure}

\begin{figure}
\centering
\includegraphics[scale=0.5]{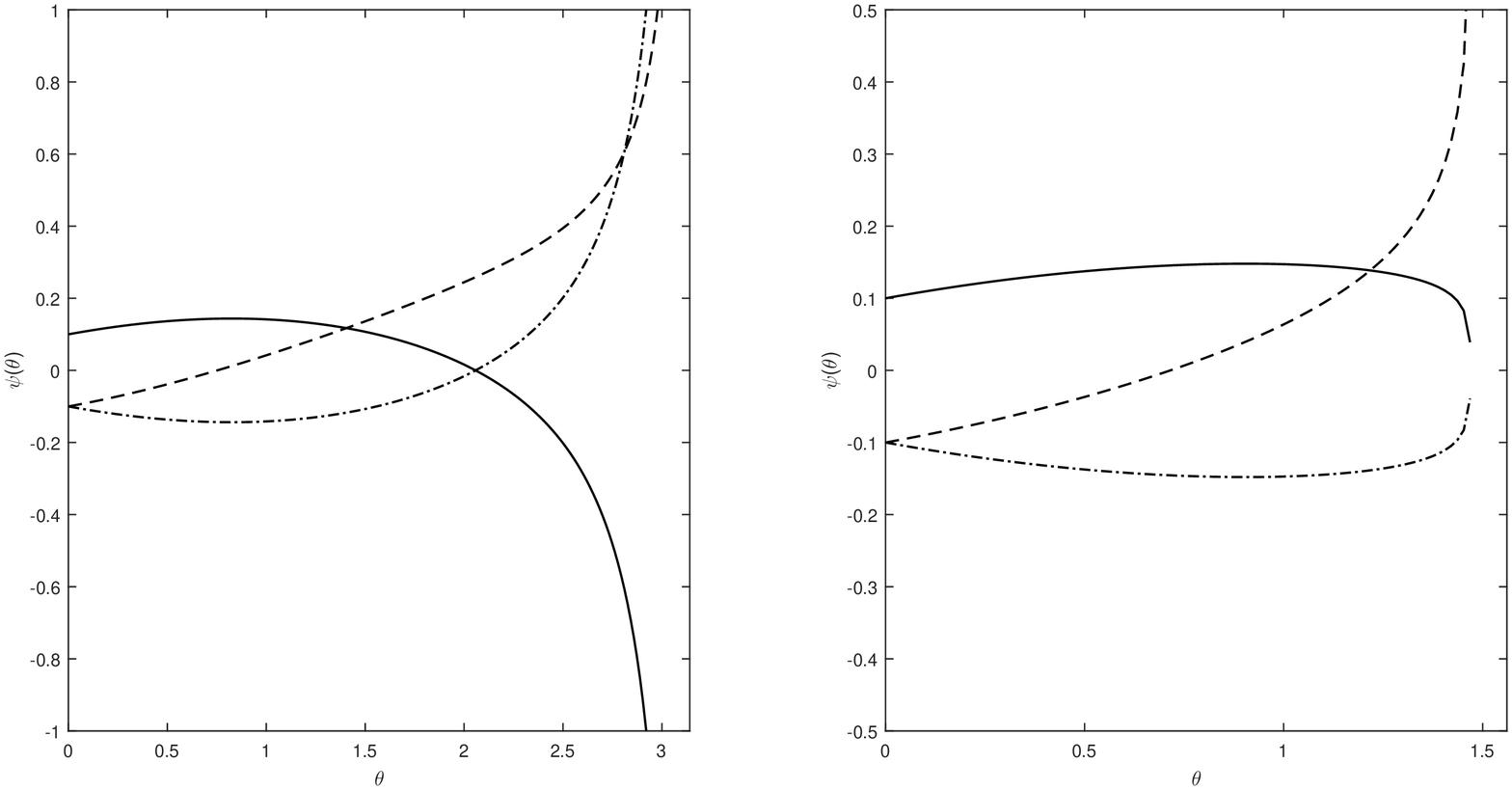}
\caption{Numerical solution of equation (\ref{e13}) for $I=-1$ (left figure) awnd $I=0.5$ (right figure). Solid line is for the initial conditions $(0.1,0.1)$, dash-dash line is for $(-0.1,0.1)$ while dash-dot line is for the initial condition $(0.1,-0.1)$}
\label{fig3}
\end{figure}

In \cite{CA} a class of dynamical systems was considered which included the
autonomous Ermakov system and the anisotropic Keplerian central force
systems and it was shown that they were linearisable up to a pair of
quadratures. In both \cite{HG} and \cite{CA} the possibility of
linearisation rested upon the existence of the LRR invariant which was
exploited for this purpose. The central feature of the demonstration
consisted in the transformation of the radial coordinate $r$ by invoking an
inverse transformation of the form $\psi=\rho(t)/r$ and in using the LRR
invariant to replace the time derivative with the derivative with respect to
the angular coordinate, $\theta$, via the relation $\dot{\theta}=h(\theta;
I)/r^2$ obtained from (\ref{e6}) with $h^2=2(I-V(\theta))$. For the
generalised Ermakov system Hass and Goedert then showed that the resulting
form of the radial equation is
\begin{equation}  \label{e7}
h^2\frac{d^2\psi}{d\theta^2}+\frac{1}{2}\frac{\partial h^2}{\partial \theta}%
\frac{d\psi}{d\theta} +(h^2+f(\theta))\psi=\frac{\rho^3(\ddot{\rho}+w^2\rho)%
}{\psi^3}.
\end{equation}
Finally to achieve linearisation they demanded that the right-hand-side be
of the form
\begin{equation}  \label{e8}
\frac{\rho^3(\ddot{\rho}+w^2\rho)}{\psi^3}=a(\theta;I)\frac{d\psi}{d\theta}%
+b(\theta;I)\psi+c(\theta;I),
\end{equation}
where $a$, $b$ and $c$ are arbitrary functions of their respective
arguments. This condition imposes the necessary restrictions upon the
frequency $w$. If one sets $\ddot{\rho}+w^2\rho=0$, then of course (\ref{e7}%
) automatically becomes linear. To illustrate the above procedure we
consider an example.\newline

\begin{equation}  \label{e9}
\ddot{r}-r\dot{\theta}^2=0,
\end{equation}
\begin{equation}  \label{e10}
r\ddot{\theta}+2\dot{r}\dot{\theta}=\frac{1}{r^3}\sin\theta.
\end{equation}
The force $F=\sin\theta/r^3 {\mathbf{e}}_\theta$ is single-valued and
periodic in the angular coordinate and is in the transverse direction. It is
clear that $\nabla\times F=-2\sin\theta/r^4$ and is nonvanishing provided $%
\theta\ne\pm\pi$. Comparison with (\ref{e5}) shows that $U(\theta)=0$ and
$V(\theta)=\cos\theta$ and the LRR invariant yields $\dot{\theta}=h^2(\theta;
I)/r^2$, where $h^2=2(I-\cos\theta)$. Under the transformation $r=1/\psi$ we
find that (\ref{e7}) reduces to
\begin{equation}  \label{e13}
\frac{d^2\psi}{d\theta^2}+\frac{\sin\theta}{(I-\cos\theta)}\frac{d\psi}{%
d\theta}+\psi=0.
\end{equation}
Some numerical solutions of the latter equation and  for $|I|>1$ are given in figures \ref{fig1} and \ref{fig2}. It is now straightforward to deduce the basis of solutions. While for $|I|<1$ in \ref{fig3}.

\noindent \textbf{Remark:} The LRR invariant is marked by an absence of any
dependence upon the radial velocity and the procedure described above relies
upon this as it enables one to solve for $\dot\theta$ in terms of $(r,\theta)
$ only.\newline

A depiction of the orbit of the motion in the plane, $(r,\,\theta) $, for
the system described by equations (2.9) and (2.10) may be obtained by the
numerical evaluation of
\begin{equation}
rr^{\prime \prime }- 2r^{\prime 2 }+ \frac {rr^{\prime }\sin\theta} {2 (I
-\cos\theta)} - r ^ 2 = 0  \label{1.1}
\end{equation}
in which the prime denotes differentiation with respect to the polar angle, $%
\theta $, or by that of (2.11). In both cases it is necessary to substitute
the value of the Lewis-Ray-Reid Invariant. We provide two orbits, one for
each option, which suggest that the latter is preferable even if
geometrically less satisfactory.
\section{Isotropic noncentral forces}

Following Berry and Shukla we assume
\begin{equation}  \label{e15}
\ddot{r}-r\dot{\theta}^2=0,
\end{equation}
\begin{equation}  \label{e16}
r\ddot{\theta}+2\dot{r}\dot{\theta}=f(r).
\end{equation}
In general it is assumed that $f(r)=C r^\mu$, where $C$ is a constant and
may be scaled to unity.  Clearly $F=f(r){\mathbf{e}}_\theta$ and the
condition $\nabla\times F\ne 0$ implies that  $\partial(r f(r))/\partial
r\ne 0$ which precludes the case $f(r)=C/r$, i.e., the $\mu=-1$  case which
was separately examined in detail in their work.  In order to map the system
(\ref{e15})-(\ref{e16}) to the Emden-Fowler equation they introduced the
transformation $(r, \theta)\longrightarrow (\tilde{T}, \tilde{J})$, where $%
\tilde{J}=r^2\dot\theta$ is the angular momentum and is not an invariant and
$\tilde{T}$ is the integrated torque defined by
\begin{equation}  \label{e17}
\tilde{T}=\int r f(r) dr.
\end{equation}
It follows from (\ref{e16}) that
\begin{equation*}
\frac{d\tilde{J}}{dt}=r f(r)
\end{equation*}
and therefore
\begin{equation}  \label{e18}
\tilde{T}^\prime(\tilde{J}) =\dot{r}.
\end{equation}
Consequently, differentiating with respect to $t$ and using (\ref{e15}) and (%
\ref{e16}),  we obtain
\begin{equation}  \label{e19}
\tilde{T}^{\prime\prime}(\tilde{J})=\frac{\tilde{J}^2}{r^4f(r)}.
\end{equation}
Thus, when $f(r)=r^\mu$, we have, on making use of (\ref{e17}) in (\ref{e19}%
),
\begin{equation}  \label{e21}
\tilde{T}^{\prime\prime}(\tilde{J})=A\tilde{J}^n\tilde{T}^m,
\end{equation}
where
\begin{equation}  \label{e22}
A=(\mu+2)^m,\;\;\; m=-\frac{\mu+4}{\mu+2},\;\;\; n=2.
\end{equation}
Eqn. (\ref{e21}) is the well-known Emden-Fowler equation \cite{PL1, PL2,
PL3, PL4}. The scaling transformations,
\begin{equation}  \label{e21a}
J=(\mu+2)^{1/n+2}\tilde{J} \quad \mbox{\rm and} \quad T=(\mu+2)\tilde{T},
\end{equation}
enable us to remove the factor $A$ in the Emden-Fowler equation which now
has the appearance
\begin{equation}  \label{e22a}
T^{\prime\prime}(J)=J^nT^m\;\;\mbox{with}\;\;n=2, m=-\frac{\mu+4}{\mu+2}.
\end{equation}

\smallskip

Extending the technique of integrable modulation we obtain \cite{GCG} the
following proposition.
\begin{prop}
The second-order ordinary differential equation \,\, $T^{\pr\pr}+d J^n T^m=0$
admits a first integral of the form
$$I=\frac{1}{2}(T^\pr J-T)^2+V(J,T)$$ where
$V(J,T)=dJ^{n+2}T^{m+1}/(m+1)$ and $n+m=-3$.
\end{prop}
Hence in our case $n=2$ implies $m= -5$.

\bigskip

Now it is known that (\ref{e22a}) admits the particular solution
\begin{equation}  \label{e23}
T(J)=\Lambda J^{(n+2)/(1-m)}, \;\;m\ne 1,
\end{equation}
where
\begin{equation}  \label{e24}
\Lambda=\left[\frac{(n+2)(n+m+1)}{(m-1)^2}\right]^{1/m-1}.
\end{equation}
Note that $m=1$ corresponds to $\mu=-3$ and it follows therefore that this
case must be separately  analysed. In fact it causes (\ref{e22a}) to reduce
to the linear equation
\begin{equation*}
T^{\prime\prime}(J)=J^2T(J)
\end{equation*}
and may also be analysed by the methods of Section 2 since it corresponds to
setting $V(\theta)=-\theta$ in(\ref{e15}) and (\ref{e16}) which admits the
first integral
\begin{equation}  \label{e26}
I=\frac{1}{2}(r^2\dot\theta)^2-\theta
\end{equation}
and leads to the linear equation
\begin{equation}  \label{e24a}
\frac{d^2\psi}{d\theta^2}+\frac{1}{(\theta+I)}\frac{d\psi}{d\theta}+\psi=0.
\end{equation}
Consequently on the level surface, $I=c$, if we set $x=\theta+c$ then the
above equation becomes
\begin{equation*}
\frac{d^2\psi}{dx^2}+\frac{1}{x}\frac{d\psi}{dx}+\psi=0.
\end{equation*}
Moreover from (\ref{e24}) we observe that, as $n=2$, so, when $m=-3$, then $%
\Lambda=0$ and the solution is trivial. However, $m=-3$ corresponds to $%
\mu=-1$ and the force $f(r)=1/r$ therefore requires special treatment.%
\newline

From (\ref{e17}) after taking into account the scaling we have
\begin{equation}  \label{e29}
T=[r^{\mu+2}-r_0^{\mu+2}],
\end{equation}
where $r_0$ is the initial position, which gives $T$ as a function of $r$
while (\ref{e23}) gives $T$ as a function of $J$, the angular momentum. By
combining these we obtain $J$ as a function of $r$ given by
\begin{equation}  \label{e30}
J=\left[\frac{1}{\Lambda}(r^{\mu+2}-r_0^{\mu+2})\right]^{\frac{1-m}{n+2}}.
\end{equation}

The equation for the orbit can be obtained using the fact the (\ref{e18})
gives
\begin{equation*}
\dot{r}=\frac{dr}{d\theta}\dot\theta=T^\prime(J)
\end{equation*}
so that
\begin{equation}  \label{e31}
\frac{d\theta}{dr}=\frac{J}{r^2 T^\prime(J)},
\end{equation}
whence using (\ref{e23}) and (\ref{e30}) we obtain the final equation
determining the orbit as
\begin{equation}  \label{e32}
\frac{d\theta}{dr}=\left(\frac{1-m}{n+2}\right)\Lambda^{-\frac{2(1-m)}{n+2}%
}(r^{\mu+2}-r_0^{\mu+2})^{-\frac{2(n+m+3)}{n+2}}.
\end{equation}
The solutions of $r$ and $\theta$ as functions of $t$ follow from (\ref{e16}%
) which upon taking into consideration the scaling of $\tilde{J}$ becomes
\begin{equation*}
(\mu+2)^{-1/n+2}\frac{dJ}{dt}=rf(r).
\end{equation*}
Scaling of the time, $t\longrightarrow\tau=(\mu+2)^{1/n+2}t$, causes it to
become $dJ/d\tau=rf(r)$ so that
\begin{equation}  \label{e33}
\tau=\int \frac{dJ}{dr}\frac{dr}{rf(r)}.
\end{equation}
Using (\ref{e30}) to calculate $dJ/dr$ we find that
\begin{equation}  \label{e34}
\tau=\left(\frac{1-m}{n+2}\right)(\mu+2)\Lambda^{\frac{n+m+1}{n+2}}\int
(r^{\mu+2}-r_0^{\mu+2})^{-\frac{n+m+3}{n+2}}dr +\tau_0
\end{equation}
which explicitly determines $\tau$ and hence $t$ as a function of $r$ and
serves in principle to determine $r=r(t)$. We then solve the equation for
the orbit, which gives $\theta=\theta(r)$, and we may recover $\theta$ as a
function of $t$ by replacing $r$. The solvable case $\mu=-4$ and the
solutions for $-1<\mu<1$ may now be easily recovered by appropriately
choosing $r_0$ as either $\infty$ or zero and they verify the corresponding
results stated in \cite{BS}.

\subsection{Symmetries and first integrals of $T^{\prime \prime} = J^2T^m$
for $m = -5,-7$}

Eqn. (\ref{e22a}) admits the Lagrangian
\begin{equation}  \label{e35}
L=T^{\prime 2}+\frac{2}{m+1}J^nT^{m+1}
\end{equation}
and corresponding Hamiltonian is given by
\begin{equation*}
H = T^{\prime 2} - \frac{2}{m+1}J^nT^{m+1}. \eqno(3.22a)
\end{equation*}

\smallskip

We examine Noether symmetries for two values of $\mu = - 3/2, - 5/3$ or $m =
-5, -7$ ( since $m = - \frac{\mu + 4}{\mu + 2}$) below:\newline

\textbf{Case a)} When $m=-5$, the Emden-Fowler equation has the form
\begin{equation*}
T^{\prime\prime}=J^2T^{-5}
\end{equation*}
and admits the symmetry generators ($G=\xi\partial_x+\eta \partial_y$),
\begin{equation}  \label{e36}
G_1=J\partial_J+\frac{2}{3}T\partial_T
\end{equation}
\begin{equation}  \label{e37}
G_2=J^2\partial_J+JT\partial_T.
\end{equation}
The Lagrangian in this case is
\begin{equation*}
L=T^{\prime 2}-\frac{J^2}{2T^4}
\end{equation*}
and from Noether's theorem it is known that the first integral is in general
given by
\begin{equation*}
I=V-\xi L -(\eta-T^\prime \xi)\frac{\partial L}{\partial T^\prime}
\end{equation*}
for some suitable function $V$.
For $G_2$ as given above we find upon setting $V=T^2$ that the Lagrangian
yields the first integral
\begin{equation}  \label{e38}
I=(JT^\prime-T)^2+\frac{J^4}{2T^4},
\end{equation}
which is identical to that derived by Berry and Shukla after reverting to
the original variables. \newline

\textbf{Case b)} In a similar fashion with $m=-7$ the corresponding
Emden-Fowler equation is derivable from the  Lagrangian
\begin{equation*}
L=T^{\prime 2}-\frac{J^2}{3T^6}.
\end{equation*}
Unlike the previous case we have only one symmetry generator which here is
given by
\begin{equation}  \label{e39}
G_1=J\partial_J+\frac{1}{2}T\partial_T,
\end{equation}
and gives rise to the first integral
\begin{equation}  \label{e40}
I=T^\prime(JT^\prime-T)+\frac{J^3}{T^6}
\end{equation}
upon setting $V=0$ and reduces to the corresponding result stated in \cite%
{BS} once we revert to  the original polar variables.

It is worth to note that Berry and Shukla considered geometric symmetries of
the force, not of the Hamiltonian or Lagrangian.

\section{Curl forces in the presence of a dissipative force}

The programme of analysis of curl forces initiated by Berry and Shukla can
be extended to curl forces in the presence of a dissipative force. Consider
a particle moving under a force ${\mathbf{F}}$ including velocity-dependent
forces given by
\begin{equation}
{\mathbf{F}}(r,\dot{r})=F_{r}(r){\mathbf{e}}_{r}+F_{\theta }{\mathbf{e}}%
_{\theta }+f_{d}(r,\dot{r}){\mathbf{e}}_{r}.
\end{equation}%
The components of the equation of motion in terms of polar coordinates $r$
and $\theta $ are
\begin{align}
\ddot{r}-r\dot{\theta}^{2}& =F_{r}(r)+f_{d}(r,\dot{r})\quad \mbox{\rm and}
\label{e101} \\
r\ddot{\theta}+2\dot{r}\dot{\theta}& =F_{\theta }(r).
\end{align}%
When we use the angular momentum, $\tilde{J}=r^{2}\dot{\theta}$, we can
express the radial equation as
\begin{equation}
\ddot{r}-\frac{{\tilde{J}}^{2}}{r^{3}}=f_{d}(r,\dot{r}).  \label{e102}
\end{equation}%
The integrated torque equation is given by
\begin{equation}
\tilde{T}^{\prime \prime }(\tilde{J})=\frac{\displaystyle{\frac{\tilde{J}^{2}%
}{r^{3}}}+f_{d}}{rF_{\theta }(r)}=\frac{\tilde{J}^{2}}{r^{4}F_{\theta }(r)}+%
\frac{f_{d}}{rF_{\theta }}.  \label{e103}
\end{equation}%
Thus, when $F_{\theta }(r)=r^{\mu }$ and $f_{d}(r,\dot{r})=r^{\nu }\dot{r}$,
we have
\begin{equation}
\tilde{T}^{\prime \prime }(\tilde{J})-\tilde{T}^{\prime }(\tilde{J})\tilde{T}%
^{\lambda }=J^{2}\tilde{T}^{\sigma },  \label{e104}
\end{equation}%
where
\begin{equation*}
\lambda =-\frac{\mu +4}{\mu +2}\quad \mbox{\rm and}\quad \sigma =\frac{\nu }{%
\mu +2}.
\end{equation*}

For $\lambda =0$ case eqn (\ref{e104}) can be identified with one of the
equation of Kamke's list. Equation (\ref{e104}) is a nonautonomous equation.
However, it can be written as the following autonomous system \cite{Aminova}%
\begin{eqnarray}
\frac{d^{2}\tilde{T}}{ds^{2}}-\tilde{J}^{2}\tilde{T}^{\sigma }\left( \frac{d%
\tilde{J}}{ds}\right) ^{2} &=&0  \label{e105} \\
\frac{d^{2}\tilde{J}}{ds^{2}}-\tilde{T}^{\lambda }\left( \frac{d\tilde{J}}{ds%
}\right) ^{2} &=&0,  \label{e106}
\end{eqnarray}%
where $s$ is a new affine parameter and now $\tilde{T}=\tilde{T}\left(
s\right) $ and $\tilde{J}=\tilde{J}\left( s\right) $. The system (\ref{e105}%
), (\ref{e106}) describes the free motion (geodesic equations) of a particle
in a two-dimensional manifold with symmetric connection coefficients
\begin{equation}
\tilde{\Gamma}_{\tilde{J}\tilde{J}}^{\tilde{T}}=-\tilde{J}^{2}\tilde{T}%
^{\sigma }~,~\tilde{\Gamma}_{\tilde{J}\tilde{J}}^{\tilde{J}}=-\tilde{T}.
\label{e106a}
\end{equation}%
Hence for the determination of the point symmetries of the system (\ref{e105}%
), (\ref{e106}) the results of \cite{mtan} can be applied.

Therefore we have that in general the system (\ref{e105}), (\ref{e106})
admits the Lie point symmetries $\tilde{G}_{1}=\partial _{s},~\tilde{G}%
_{2}=s\partial _{s}$. However, when~$\sigma =1+4\lambda $, i.e. $\nu =-3\mu
-14,$ there exists the extra Lie point symmetry%
\begin{equation}
\tilde{G}_{3}={\lambda }\tilde{J}\partial _{\tilde{J}}-\tilde{T}\partial _{%
\tilde{T}}.
\end{equation}

Moreover $\tilde{G}_{3}$ is also the unique Lie point symmetry of equation (%
\ref{e104}). The zeroth- and the first-order invariants of $\tilde{G}_{3}$
are%
\begin{equation}
w=\left( \tilde{J}\right) ^{\frac{1}{\lambda }}\tilde{T}~,~u=\left( \tilde{J}%
\right) ^{\frac{\lambda +1}{\lambda }}\tilde{T}^{\prime }.
\end{equation}

We select $w$ to be the new independent variable and $u=u\left( w\right) $
to be the new dependent variable. Therefore in the new variables, $\left\{
w,u\left( w\right) \right\} $, equation (\ref{e104}) is reduced to the
following first-order equation%
\begin{equation}
\left( \lambda u+w\right) \frac{du}{dw}=\left( 1+\lambda \left( 1+w^{\lambda
}\right) \right) u+\lambda w^{1+4\lambda }\text{.}  \label{e107}
\end{equation}%
which is an Abel's Equation of the second type \cite{Bougoffa}.

Furthermore, when $\lambda =0,~\sigma =1$, i.e. $\mu =-4$, $\nu -2\,$,
equation (\ref{e104}) admits eight Lie point symmetries, which means that
there exists a transformation $\left\{ \tilde{J},\tilde{T}\left( \tilde{J}%
\right) \right\} \rightarrow \left\{ X,Y\left( X\right) \right\} $ and (\ref%
{e104}) becomes $\frac{d^{2}Y}{dX^{2}}=0$.

On the other hand we introduce the variables $\left\{ J,T\left( J\right)
\right\} \rightarrow \left\{ Y\left( z\right) ,\frac{dY\left( X\right) }{dz}%
\right\} $, where the nonautonomous second-order equation (\ref{e104})
becomes the autonomous third-order equation
\begin{equation}
\left( \frac{dY}{dz}\right) \left( \frac{d^{3}Y}{dz^{3}}\right) -\left(
\left( \frac{d^{2}Y}{dz^{2}}\right) +\left( \frac{dY}{dz}\right) ^{2+\lambda
}\right) \left( \frac{d^{2}Y}{dz^{2}}\right) -Y^{2}\left( \frac{dY}{dz}%
\right) ^{\sigma +3}=0,  \label{e108}
\end{equation}%
which admits always the symmetry vector $Z^{1}=\partial _{z}$. Reduction
with the latter symmetry vector leads to (\ref{e104}). However for specific
values of the constants $\lambda ,\sigma $, i.e. $\mu ,\nu $, equation (\ref{e108}) admits extra symmetry vectors.

Specifically we have that for arbitrary $\lambda ,\sigma $ , \ the admitted
Lie symmetry is the $Z^{1}$. For $\lambda =0$, and $\sigma =1,$ equation (%
\ref{e108}) admits the Lie symmetries $Z^{1}=\partial _{z}$, \ $%
Z^{2}=z\partial _{z}$. For $\lambda =-1$, and $\sigma =-3$, equation (\ref%
{e108}) admits the Lie symmetries $Z^{1}$ and $Z^{3}=Y\partial _{Y}$, while
for ~$\sigma =1+4\lambda ,\lambda \neq -1$ we have that  (\ref{e108}) is
invariant under the two dimensional Lie algebra$~\left\{ Z^{1},Z^{4}\right\}
$ in which
\begin{equation}
Z^{4}=\left( 1+\lambda \right) z\partial _{z}+\lambda Y\partial _{Y}
\label{e109}
\end{equation}%
this vector field is related with vector field $\tilde{G}_{3}$ of above.

Again from (\ref{e109}) we can see that $Y\left( z\right) =Y_{0}z^{\frac{%
\lambda }{1+\lambda }}$, is a solution of (\ref{e108}) if and only if
\begin{equation}
\left( \lambda \right) ^{-2}\left( \lambda Y_{0}\right) ^{4+4\lambda
}-\left( \lambda Y_{0}\right) ^{1+\lambda }\left( 1+\lambda \right)
^{3\lambda }-\left( 1+\lambda \right) ^{1+4\lambda }=0,
\end{equation}%
while when $\lambda =-1$, a special solution is the exponential function $%
Y\left( z\right) =Y_{0}\exp \left( -z\right) .$

The general analysis of equation (\ref{e104}) is of interests however that
is overpass the purpose of this current work and will be published in a
forthcoming paper.

\section{Outlook and discussion}

We have carried forward the programme of analysis of curl forces initiated
by Berry and Shukla. This is largely an unexplored area of nonlinear
mechanics, though efforts at linearisation of planar systems subject to
nonisotropic central forces were performed by Athorne and Haas and Goedert
in the context of Kepler-Ermakov theory. We have shown that the first
integrals derived by Berry and Shukla are the Noetherian first integrals
resulting from the symmetries of the Emden-Fowler equation.

\section*{Acknowledgements}

We wish to thank Sir Michael Berry
for various comments and suggestions that have been helpful to
improve the manuscript. We are greateful to Pragya Shukla and Pepin Cari\~{n}%
ena for enlighting discussions and constant encouragement. A.P. acknowledges financial
support \ of FONDECYT grant no. 3160121.


\begin{thebibliography}{99}
\bibitem{Aminova} AV Aminova \& NAM Aminov, \emph{The projective geometric
theory of systems of second-order differential equations: straightening and
symmetry theorems}, Sb Math 201 (2010) 631

\bibitem{CA} C Athorne, \emph{Kepler-Ermakov problems} J Phys A: Math Gen 24
(1991) L1385-L1389

\bibitem{BS} MV Berry \& Pragya Shukla, \emph{Classical dynamics with curl
forces, and motion driven by time-dependent flux} J Phys A: Math Theor 45
(2012) 305201 (18pp)

\bibitem{BS1} MV Berry \& Pragya Shukla, \emph{Physical curl forces: dipole
dynamics near optical vortices}, J Phys A: Math Theor 46 (2013) 422001 (9pp)

\bibitem{BS2} MV Berry \& Pragya Shukla, \emph{Hamiltonian curl forces},
submitted to Proc R Soc A

\bibitem{Bougoffa} L Bougoffa, \emph{New exact general solutions of Abel
equation of the second kind}, Appl Math Comput 216, 689 (2010)

\bibitem{Chandra} S Chandrasekhar, \emph{An Introduction to the Study of
Stellar Structure}, Dover Publications Inc, New York (1957)

\bibitem{Emden} R Emden, \emph{Gaskugeln, Anwendungen der mechanischen
Warmen-theorie auf Kosmologie und meteorologische Probleme} (Leipzig,
Teubner, 1907)

\bibitem{Forsyth} AR Forsyth, \emph{A Treatise of Differential Equations},
MacMillan \& Co, Vol l (London) (1956)

\bibitem{GL} VM Gorringe \& PGL Leach, \emph{Conserved vectors for the
autonomous system $r^{\prime \prime} + g(r,\theta)\hat{r} + h(r,\theta)\hat{%
\theta} =0$} Phys D 27 (1987), no. 1-2, 243-248

\bibitem{PL4} KS Govinder \& PGL Leach, \emph{Integrability analysis of the
Emden-Fowler equation} J Nonlin Math Phys (2007) 14 435-453

\bibitem{GCG} P Guha \& A Ghose Choudhury, \emph{Integrable Time-Dependent
Dynamical Systems: Generalized Ermakov-Pinney and Emden-Fowler Equations},
Nonlin Dynam Syst Theor, 14 (4) (2014) 355--370

\bibitem{HG} F Haas \& J Goedert, \emph{On the linearization of the
generalized Ermakov system} J Phys A: Math Gen 32 (1999) 2835-2844

\bibitem{Lane} JH Lane, \emph{On the Theoretical Temperature of the Sun
under the Hypothesis of a Gaseous Mass Maintaining its Volume by its
Internal Heat and Depending on the Laws of Gases Known to Terrestrial
Experiment}, The American Journal of Science and Arts 50 (1870) 57-74

\bibitem{PL1} PGL Leach, \emph{First integrals for the modified Emden
equation $\ddot{q} + \alpha(t) \dot{q} + q^n = 0 $} J Math Phys (1985) 26
2510-2514

\bibitem{PL2} PGL Leach, \emph{Group theoretical treatment of the
generalized Emden-Fowler equation} Proc XVIIIth Inter Colloq on Group
Theoret Methods in Physics, VV Dodonov and VI Man'ko edd (Nova Science
Publishers, New York, 1992)

\bibitem{PL3} PGL Leach, R Maartens \& SD Maharaj, \emph{Self-similar
solutions of the generalized Emden-Fowler equation} Int J Nonlinear
Mechanics 27(4) 575-582 (1992)

\bibitem{Mach} P Mach, \emph{All solutions of the $n=5$ Lane-Emden equation}%
, J Math Phys (2012) 53 062503

\bibitem{mtan} M Tsamparlis \& A Paliathanasis, \emph{Lie and Noether
symmetries of geodesic equations and collineations}, Gen Rel Grav 42, 2957
(2010)

\bibitem{Wang 76 a} JSW Wang, \emph{On the generalized Emden-Fowler equation
} SIAM Rev. 17 (1975)

\bibitem{Whittaker 44 a} ET Whittaker, \emph{A Treatise on the Analytical
Dynamics of Particles and Rigid Bodies} (Dover, New York, 1944)
\end{thebibliography}
\end{document}